# SkyMouse: A smart interface for astronomical on-line resources and services

Chen-Zhou CUI[1], Hua-Ping SUN[1], Yong-Heng ZHAO[1], Yu LUO[1], Da-Zhi QI[2]

1. National Astronomical Observatories, Chinese Academy of Sciences, Beijing 100012, China

2. Tianjin University, Tianjin 300072, China

Correspondence to: ccz@bao.ac.cn

**Abstract**

With the development of network and the World Wide Web (WWW), the Internet has been growing and changing dramatically. More and more on-line database systems and different kinds of services are available for astronomy research. How to help users find their way through the jungle of information services becomes an important challenge. Although astronomers have been aware of the importance of interoperability and introduced the concept of Virtual Observatory as a uniform environment for future astronomical on-line resources and services, transparent access to heterogeneous on-line information is still difficult.

SkyMouse is a lightweight interface for distributed astronomical on-line resources and services, which is designed and developed by us, i.e., Chinese Virtual Observatory Project. Taking advantage of screen word-capturing technology, different kinds of information systems can be queried through simple mouse actions, and results are returned in a uniform web page. SkyMouse is an easy to use application, aiming to show basic information or to create a comprehensive overview of a specific astronomical object.

In this paper current status of on-line resources and services access is reviewed; system architecture, features and functions of SkyMouse are described; challenges for intelligent interface for on-line astronomical resources and services are discussed.

**Introduction**

With the development of network and the World Wide Web (WWW), the Internet has been growing and changing dramatically. Nowadays, it has been an indispensable part of our daily lives and work. However with the expansion of network scale and its contents, it becomes more and more difficult or even impossible to look for a specific webpage and website manually. How to help users find their way through the jungle of information services becomes an important challenge, which has been raised since the early development of the WWW (Egret, 1994). Driven by the requirement, search engine systems appeared and have been affecting the usage of the Internet deeper and deeper. For example, Yahoo![1] and Google[2] are representatives for the 2nd and the 3rd generations of web search engines, respectively. It is hard to imagine that one does not know of or does not use these engines. As one of the most powerful and popular search engine systems, Google is used by many astronomers everyday to look for bibliographies, data and many others.

---

[1] http://www.yahoo.com
[2] http://www.google.com

Although functions of these web search engines are powerful, intrinsic limitations are unsurpassable. The Internet information can be divided into three categories by its access limitation: white network, gray network, and black network. White network mainly refers to public accessible, static web pages and files. Contents of gray network are usually created on-the-fly based on backend databases. Black network refers to private and local network, which is only accessible for its given users in a specific scope. With the help of "spiders" and "robots", general web search engines can only collect information from white network and a part of gray network. Information in black network is not accessible by spiders and robots.

For professional astronomy research, results provided by these common web search engines are far from enough. In order to meet the requirements of astronomers for easy access to bibliographies and all kinds of data, which usually exist as gray or black networks, professional astronomical information systems are developed (Genova et al., 2000; Egret et al., 2000). ADS[3] (Kurtz et al, 2000), for example, is one of the best well known bibliography systems. arXiv[4] (Ginsparg 1996) is also well known by astronomers as an e-print service for its astrophysics (i.e., "astro-ph") category. SIMBAD[5] (Wenger et al. 2000) acts as a leading reference database for astronomical objects. VizieR[6] (Ochsenbein et al. 2000) and NED[7] (Madore et al. 1992) are two important catalog database systems for inner- and extra-Galactic objects, respectively. Furthermore, there are many other astronomical database systems, for example, NASA MAST[8], CADC[9], BADC[10], and so on.

Usually, in order to collect enough materials for a given research project, different astronomical information systems have to be browsed and queried. Federating results from different sources are becoming more and more common with the increasing multi-waveband research. Therefore, Virtual Observatory (VO) (Szalay et al. 2001, Brunner et al. 2001) was introduced and the International Virtual Observatory Alliance (IVOA) (Quinn et al, 2004) was established. VO aims to build a cyber-infrastructure for online astronomical research. Through defining and advocating a set of standard formats and interfaces, interoperability among different services hopes to be achieved.

Uniform access to global astronomical database systems is a basic goal for the VO. DataScope[11] from US National Virtual Observatory (NVO) Project and AstroScope[12] from UK AstroGrid Project are two leading efforts on uniform discovery and access of online astronomy resources and services. By querying of the VO Registry (Plante et al. 2004), the two systems try to discover and fetch as much data as possible for a given object or a given sky area from world wide database systems.

Chinese Virtual Observatory (China-VO) (Cui et al. 2004) is the national VO project in China initiated in 2002 by Chinese astronomical community led by National Astronomical Observatories, Chinese Academy of Sciences. The China-VO aims to provide VO infrastructure for Chinese astronomers. It focuses its research and development on VO science and applications. Based on grid middleware OGSA-DAI, the China-VO team is developing a system named VO-DAS (VO Data Access System) (Liu et al. 2007) to provide a uniform access interface for catalog, spectra, images, etc.

The SkyMouse (Cui et al. 2006) is another effort from the China-VO team in the scope of data access but in a different way, which is a project supported by the National Natural Science Foundation of China through grant No. 60603057. Taking advantage of screen word-capturing technology, different kinds of information systems including those mentioned earlier can be queried through simple mouse actions, and results can be returned in a uniform web page. Deviating from DataScope and AstroScope, the SkyMouse is a lightweight interface for astronomical on-line resources and services, aiming to show basic information or to create a comprehensive overview for a specific astronomical object.

In the second section, architecture of the SkyMouse system is described. Characters and functions of the system are introduced in the third section. Challenges for intelligent interface for on-line astronomical resources and services are discussed in the final section.

## System architecture

Inspired by Kingsoft PowerWord[13] and CiHu[14], the idea of SkyMouse was introduced by us in May 2005. Kingsoft PowerWord is a popular electrical dictionary production, esp. in China. CiHu is an application system developed by Institute of Remote Sensing Applications, Chinese Academy of Sciences to access online Geographic Information Systems (GIS) and related resources through cursor actions.

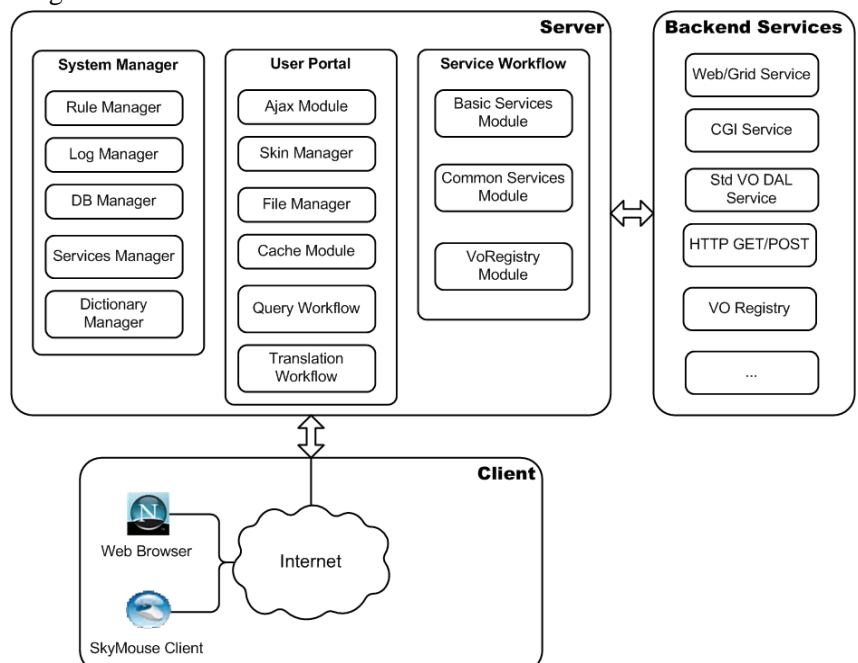

Fig. 1    B/S and C/S Architecture of the SkyMouse

A mixed architecture of Client/Server (C/S) and Browser/Server (B/S) adopted by the SkyMouse is shown in Fig. 1. The whole system is composed of three parts: client, server and backend services.

Client of the system can be a commonly used web browser or a SkyMouse client, a package developed specially for the system. To provide an easy-to-use interface, a dedicated screen word-capturing client is designed and developed for both Microsoft Windows OS[15] and Linux OS[16]. It is a thin client design. Core functions of the client include capturing and identifying words displayed on the screen by a cursor, and then sending them to the SkyMouse server. When the results are returned by the server, a floating window displays these results.

Server is the core of the system. It includes three sub-systems: system manager, user portal and service workflow. SkyMouse server receives a phrase from the client, submits it as a query to the backend resources and services; integrates returned query results into a uniform page and feeds it back to the client. Main functions of system manager include managements of system rules, logging, server side database, service information, server side dictionary and others. User portal sub-system provides management functions for AJAX (Garrett 2005) modular, page layout (i.e. Skin), cache, multimedia files, query workflow, translation workflow, etc. Service workflow sub-system is the interface for backend heterogeneous services, providing interfaces to CGI, HTTP GET/POST, Web/Grid services, etc. VO Registry client modular is also provided here. Service workflow sub-system is in charge of workflow among services. Several cutting-edge technologies are adopted here, including AJAX, Web/Grid services, .Net, and VO technologies (Sun et al. 2007).

Backend resources and services for the SkyMouse are mainly contributed by third parties. Many kinds of current online astronomical information systems and services, including CGI scripts, HTTP GET applications, Web/Grid services, can be accessed and queried by the SkyMouse. Several well-known astronomical online systems, such as SIMBAD, VizieR, NED, ADS, astro-ph@arXiv, are selected as default basic services. For those basic services, optimized query client codes are developed. Additional systems or services can be added easily to the system by administrator and users. VO Registry based service search and discovery are supported by the SkyMouse. An intelligent common interface is developed for these additional services.

## Features and functions

A unique feature of the system is its use mode. With the help of a dedicated SkyMouse client, SkyMouse provides a completely new access mode of astronomical online resources and services, which makes information gathering easier than previous attempts. A typical scenario is described as follows.

An astronomer meets an unacquainted object name when he is reading a paper online, so he moves the cursor on to the word(s). The word(s) will be captured and identified automatically by SkyMouse client. A small floating window appears. If the phrase under the cursor is an astronomical object, SIMBAD and NED database systems will be queried automatically. Basic information, including coordinates (RA, DEC), magnitude, type, distance, etc. will appear on the floating window. Fig. 2 is an example for "M31". If the phrase is not an astronomical object, the online dictionary system will be queried and a brief interpretation of it will appear on the window. If the astronomer needs more information than those appeared on the floating window, he can click shortcut buttons on the window to query more astronomical information systems or submit the phrase to Google. If he selects searching by SkyMouse, then bibliographies, images, observations and much more query results from pre-selected information systems will be displayed as a webpage in the default web browser of the astronomer's computer. If he hopes to use these results in the future, he can save them in different formats, such as HTML, PDF, TEXT, etc.

---

[15] http://www.microsoft.com
[16] http://www.linux.org

If the captured words are in Chinese, SkyMouse will translate them into English first and then continue working as mentioned earlier.

SkyMouse client is also a very easy-to-use package, which can be downloaded free from the official website of the SkyMouse[17]. Unpacking is the only required action before its execution.

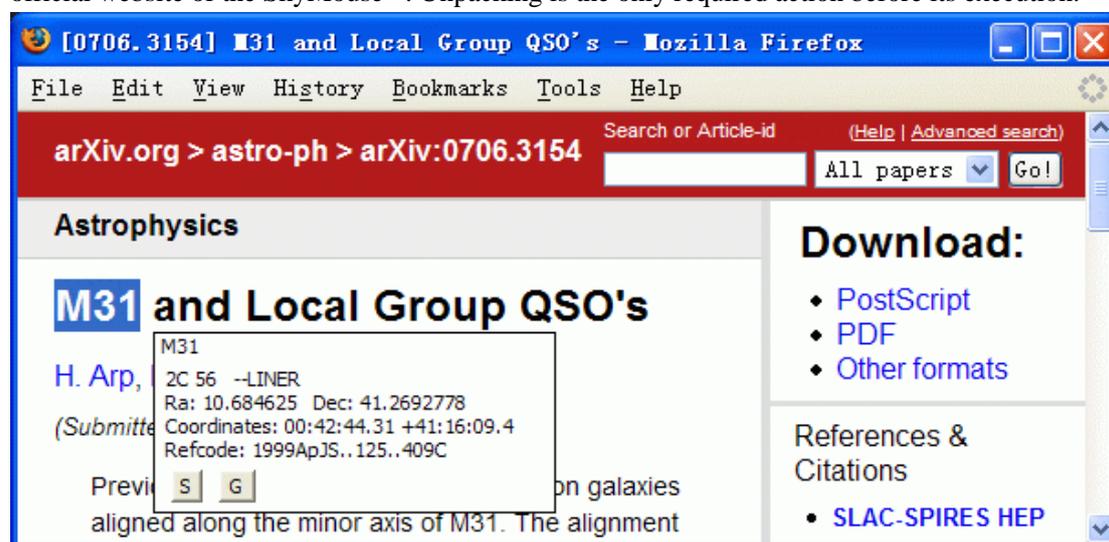

Fig.2    Basic information for M31 displayed on the SkyMouse client window

Another mode of the system is to submit a query to the SkyMouse website directly as browsing usual web pages.

From the above introduction, we can see that SkyMouse can be regarded as a combination of online astronomical dictionary and an integrated client interface for on-line astronomical resources and services. SkyMouse 1.0, the first public version, was released in July 2006. SkyMouse 2.0 is the latest version and was released in June 2007. Main features of this version include:

- Intelligent screen word-capturing client for both Windows and Linux operation systems with cursor and select word-capturing modes
- Multiple types of backend services support, for example HTTP GET, CGI, Web/Grid Services
- Pre-selected commonly used services, including SIMBAD, NED, DSS, ADS, VizieR and astro-ph@arXiv
- Multiple output formats, i.e., PDF, HTML, ASCII, etc.
- VO-Registry based service search and discovery
- Easy information management of backend services and resources

- Increased user-friendliness due to AJAX technology
- Prompt responses based on caching
- Chinese to English translation.

More information and documents can be found at the SkyMouse website "http://skymouse.china-vo.org".

# Conclusions

Taking advantage of a dedicated screen word-capturing client, SkyMouse is designed and developed to provide a unique and simple interface for VO services and other astronomical on-line resources. Through simple cursor movement, world-wide astronomical databases and services can be queried. Users don't need to worry about the underlying backend technology.

Although basic goals for the system are "easy of use" and "flexibility", content merging poses a huge problem to the system. It is an important challenge that researchers in the fields of computer science and artificial intelligence must deal with effectively. A growing number of astronomical resources and services are made available through the Internet. Responses generated by submission of queries to a set of heterogeneous resources are difficult to merge or integrate, because different providers generally use different data formats and metadata descriptions. That is why "interoperability" becomes a basic goal of the IVOA. At this present stage, we have to leave the challenge to others.

Another challenge for the system is intelligent identification of captured words. How to send a meaningful or user requested phrase to the server is a difficult semantic question. A native solution is a dictionary based word coupling. But it is only a partial solution, and will not work all the time. Therefore, besides automatic cursor word-capturing mode, user selection mode is designed for SkyMouse client. Through a "selection" action, precise phrase can be captured and then submitted.

Although there are many challenges, the direction of uniform access to distributed heterogeneous on-line resources and services is an important one and must not be ignored. With the maturing and wide adopting of IVOA specifications and recommendations, more and more VO-ready resources and services will come up. Smart user interfaces such as the SkyMouse must be welcomed by astronomers and other users.

*Acknowledgments:* This work is funded by the National Natural Science Foundation of China through grant No. 60603057, 90412016 and 10778623. All the China-VO team members are acknowledged for their tightly collaboration. LAMOST project provides the necessary work environments for the China-VO. Cui C. thanks Jayant Gupchup, Johns Hopkins University, for proof-reading the paper carefully.